\documentclass[twocolumn,tighten]{aastex62}
\accepted{2018 August 28}
\received{2018 June 4}
\submitjournal{AJ}
\usepackage{graphicx}
\usepackage{natbib}
\usepackage{latexsym}
\usepackage{amssymb}
\usepackage{longtable}
\usepackage{amsmath}
\usepackage{url}
\def\msun{\ifmmode {\rm\,M_\odot}\else ${\rm\,M_\odot}$\fi}
\def\Msun{\ifmmode {\rm\,\it{M_\odot}}\else ${\rm\,M_\odot}$\fi}
\def\lsun{\ifmmode {\rm\,L_\odot}\else ${\rm\,L_\odot}$\fi}
\def\Lsun{\ifmmode {\rm\,\it{L_\odot}}\else ${\rm\,L_\odot}$\fi}
\def\rsun{\ifmmode {\rm\,R_\odot}\else ${\rm\,R_\odot}$\fi}
\def\Rsun{\ifmmode {\rm\,\it{R_\odot}}\else ${\rm\,R_\odot}$\fi}
\def\Tsun{\ifmmode {\rm\,T_\odot}\else ${\rm\,T_\odot}$\fi}
\def\arcsec{\ifmmode {^{\prime\prime}}\else $^{\prime\prime}$\fi}
\def\asec{\ifmmode {^{\prime\prime}}\else $^{\prime\prime}$\fi}
\def\arcmin{\ifmmode {^{\prime}}\else $^{\prime}$\fi}
\def\amin{\ifmmode {^{\prime}}\else $^{\prime}$\fi}
\def\simlt{\mathrel{\spose{\lower 3pt\hbox{$\mathchar"218$}}
     \raise 2.0pt\hbox{$\mathchar"13C$}}}
\def\simgt{\mathrel{\spose{\lower 3pt\hbox{$\mathchar"218$}}
\     \raise 2.0pt\hbox{$\mathchar"13E$}}}

\begin{document}

\author{P. Wilson Cauley}
\affiliation{Arizona State University, School of Earth and Space Exploration, Tempe, AZ 85287}

\author{Christoph Kuckein}
\affiliation{Leibniz-Institut f{\"u}r Astrophysik Potsdam (AIP), An der Sternwarte 16, 14482 Potsdam, Germany}

\author{Seth Redfield}
\affiliation{Wesleyan University, Astronomy Department, Van Vleck Observatory, Middletown, CT}

\author{Evgenya L. Shkolnik}
\affiliation{Arizona State University, School of Earth and Space Exploration, Tempe, AZ 85287}

\author{Carsten Denker}
\affiliation{Leibniz-Institut f{\"u}r Astrophysik Potsdam (AIP), An der Sternwarte 16, 14482 Potsdam, Germany}

\author{Joe Llama}
\affiliation{Lowell Observatory, Flagstaff, AZ 86001}

\author{Meetu Verma}
\affiliation{Leibniz-Institut f{\"u}r Astrophysik Potsdam (AIP), An der Sternwarte 16, 14482 Potsdam, Germany}

\correspondingauthor{P. Wilson Cauley}
\email{pwcauley@gmail.com}

\title{The effects of stellar activity on optical high-resolution exoplanet transmission spectra}

\begin{abstract} 

Chromospherically sensitive atomic lines display different spectra in stellar
active regions, spots, and the photosphere, raising the possibility that
exoplanet transmission spectra are contaminated by the contrast between various
portions of the stellar disk. To explore this effect, we performed transit
simulations of G and K-type stars for the spectral lines \ion{Ca}{2} K at 3933
\AA, \ion{Na}{1} 5890 \AA, \ion{H}{1} 6563 \AA\ (H$\alpha$), and \ion{He}{1}
10830 \AA. We find that strong facular emission and large coverage fractions
can contribute a non-negligible amount to transmission spectra, especially for
H$\alpha$, \ion{Ca}{2} K, and \ion{Na}{1} D, while spots and filaments are
comparatively unimportant. The amount of contamination depends strongly on the
location of the active regions and the intrinsic emission strength. In
particular, active regions must be concentrated along the transit chord in
order to produce a consistent in-transit signal.  Mean absorption signatures in
\ion{Na}{1} and H$\alpha$ for example, can reach $\approx 0.2\%$ and 0.3\%,
respectively, for transits of active latitudes with line emission similar in
strength to moderate solar flares. Transmission spectra of planets transiting
active stars, such as HD 189733, are likely contaminated by the contrast
effect, although the tight constraints on active region geometry and emission
strength make it unlikely that consistent in-transit signatures are due
entirely to the contrast effect.  \ion{He}{1} 10830 \AA\ is not strongly
affected and absorption signatures are likely diluted, rather than enhanced, by
stellar activity. \ion{He}{1} 10830 \AA\ should thus be considered a priority
for probing extended atmospheres, even in the case of active stars.

\end{abstract}


\section{INTRODUCTION}
\label{sec:intro}

Transmission spectroscopy has been the workhorse for measuring the physical
properties of exoplanet atmospheres, probing from the deep molecular layers of
hot planets \citep{sing16} out to the thermosphere
\citep{redfield08,wyttenbach15} and beyond to the unbound exosphere
\citep{vidal03,ehren15}. These observations have revealed a variety of
molecular \citep[e.g.,][]{knutson07,snellen10,deming13,kreidberg15,brogi16} and
atomic species in exoplanet atmospheres
\citep[e.g.,][]{jensen11,pont13,cauley15,wilson15,wyttenbach17,casasayas17,spake18},
detailed the presence of cloud layers \citep{kreidberg14}, and even probed the
dynamics of bound and unbound material \citep{bourrier15,louden15,brogi16}.
The \textit{James Webb Space Telescope} and upcoming extremely
large ground-based telescopes will greatly expand the sample of planetary
atmospheres that can be recorded in transmission.

Exoplanet transmission spectra are always contaminated by star spots and
faculae to some degree: the stellar disk is heterogeneous and the weighting of
the integrated in-transit signal towards the unocculted regions of the stellar
disk can produce spurious features in the transmission spectrum
\citep{berta11,sing11,cauley17a,rackham18}. We hereon refer to this phenomenon
as the \textit{contrast effect} due to its origin in the stark differences
between the spectra of various active region features.  Time-variable levels of
stellar activity, as opposed to the effects induced on the observed spectrum by
the planet's shadow, can also impact transmission spectra created by combining
data from different epochs \citep{zellem17}.  While more attention has recently
been paid to the magnitude of the contrast effect, which, for example, has been
used to estimate faculae coverage for GJ 1214 \citep{rackham17} and possible
contamination of H$\alpha$ transmission spectra for HD 189733 b
\citep{cauley17a}, a more complete understanding is needed of the magnitude of
contrast contamination in exoplanet transmission spectra.

%Sampling the stellar
%surface occulted by the planet as a function of transit time can similarly be
%used to derive precise 3D stellar obliquities, an important ingredient in
%validating planetary migration theories \citep{cegla16}.

Progress on this front has been made by \citet{rackham18}, who showed how
active regions and spots on M-dwarfs can affect the strength of molecular
features in the transmission spectra of small rocky planets. For many faculae
and spot configurations, important molecular features, such as O$_2$, H$_2$O,
and CO$_2$, can be contaminated at the level of $\approx 10-30\%$. These
results highlight the need for stronger constraints on facular and spot
coverage fractions in order to interpret transmission spectra of transiting
planets around M-dwarfs.

While molecular features such as H$_2$O and CO are observed at pressures of
$\approx 1$ bar in exoplanet atmospheres \citep[e.g.,][]{kreidberg15}, the
cores of atomic lines such as \ion{Na}{1} D, H$\alpha$, and \ion{He}{1} 10830
\AA\ sample pressures of $\approx 1$  $\mu$bar at higher altitudes
\citep{huang17,oklopcic18}. Atomic transitions were the first detections in
exoplanet atmospheres \citep{charbonneau02,redfield08,snellen08} and continue
to be important diagnostics of the thermosphere
\citep[e.g.,][]{cauley16,barnes16,cauley17a,cauley17b,chen17,khalafinejad17,wyttenbach17,casasayas17}.
Recently, \citet{spake18} reported the first detection of helium in the
extended atmosphere of WASP-107 b, confirming the potential of the \ion{He}{1}
10830 \AA\ line as a probe of hot planet exospheres, as suggested by early
theoretical studies in atmospheric characterization \citep{seager00}.
High-resolution observations of atomic lines can also provide information on
velocity flows in the atmosphere
\citep[e.g.,][]{wyttenbach15,louden15,cauley17a} and cross-correlation analysis
of high-resolution molecular absorption can yield constraints on planetary
rotation \citep{snellen10,snellen14,brogi16}. If non-negligible for hot Jupiter
systems, the contrast effect can contaminate the magnitude of the measured
absorption and may bias cross-correlation signatures towards active region
features.

In this paper we explore the effect of faculae and spots on the strength of
observed atomic absorption features in exoplanet transmission spectra. We focus
on G and K stars and spectral lines which have a significant contribution from
the chromosphere, specifically the transitions of \ion{Ca}{2} K at 3933 \AA,
the \ion{Na}{1} 5896 \AA\ component of the \ion{Na}{1} D doublet, \ion{H}{1}
6563 \AA\ (H$\alpha$), and \ion{He}{1} 10830 \AA. The details of the
simulations are outlined in \autoref{sec:model} and the model results are given
in \autoref{sec:results}.  Included in \autoref{sec:results} is a transit of
the resolved solar disk in order to provide some context for the rest of the
simulation parameters.  A discussion of how the results compare with
observations is presented in \autoref{sec:discussion} and a brief summary of
our conclusions is given in \autoref{sec:conclusion}.

\section{Simulations of active stellar surface transits}
\label{sec:model}

Our approach to simulating planetary transits of active stellar surfaces is
similar to the procedure described in \citet{cauley17a}. Each simulation
samples the integrated stellar spectrum at ten different points across the
planet's transit. The mean in-transit spectrum changes negligibly if the number
of sampled transit points is increased beyond ten. The in-transit spectrum is
then divided by the integrated out-of-transit spectrum to produce the contrast
spectrum, which we label $S_\text{in}/S_\text{out}$. $S_\text{in}/S_\text{out}$
varies depending on which portion of the stellar disk is occulted by the
planet. We note that the weaker doublet members \ion{Ca}{2} H and \ion{Na}{1} 5896 \AA\ 
are not shown here but, due to their lower oscillator strengths, their contrast signals 
are constrained to be lower than those for \ion{Ca}{2} K and \ion{Na}{1} 5890 \AA. 
The integrated model spectra are convolved with a Gaussian kernel to a resolving power of
$R \approx 70,000$ to approximate a typical high-resolution optical spectrograph.

We explore four different effective temperatures, $T_\text{eff} = 4500$, 5000,
5500, and 6000 K, with log$g = 4.5$ and [Fe/H]$= 0.0$, which covers mid-K to
early G-type main sequence stars. Two different values, 0.01 and 0.02, of the
ratio ($R_\text{p} / R_*$)$^2$ are tested. These quantities cover a majority of
the transit depths currently probed by high-resolution transmission
spectroscopy. The stellar equatorial rotational velocity is set at 2 km
s$^{-1}$ for all cases, including the solar test case, and we assume rigid
rotation. Although G and K stars shows a range of $v$sin$i$ values, the vast
majority of transiting planet hosts with $T_\text{eff} \leq 6000$ K have
$v$sin$i$ $\lesssim 6$ km s$^{-1}$. The exact value of the stellar rotational
velocity contributes significantly less to the shape of the contrast spectrum
compared with the active region parameters. For this reason a constant
$v$sin$i$ is chosen for the sake of simplifying the parameter space. The
transit impact parameter is set at $b=0.5$ for all simulations. The stellar
disk has a spatial resolution of 0.0025 $R_*$, resulting in an $800 \times 800$
Cartesian grid. The planet is simulated in an aligned prograde orbit in
all cases.  

The model spectra are generated using \texttt{SPECTRUM} \citep{gray94}. We calculate
the spectra at twenty four different values of $\mu = \cos{\theta}$, where
$\theta$ is the angle between the normal vector to the stellar surface and the
line of sight, from $\mu = 0.01 - 1.0$.  We then interpolate the ratio of the
spectra at each wavelength onto a $\mu$-grid representing the stellar disk,
resulting in a cube of $I(\mu)/I(\mu=1)$ as a function of wavelength.

\begin{figure*}[htbp]
   \centering
   \includegraphics[scale=.75,clip,trim=13mm 60mm 5mm 85mm,angle=0]{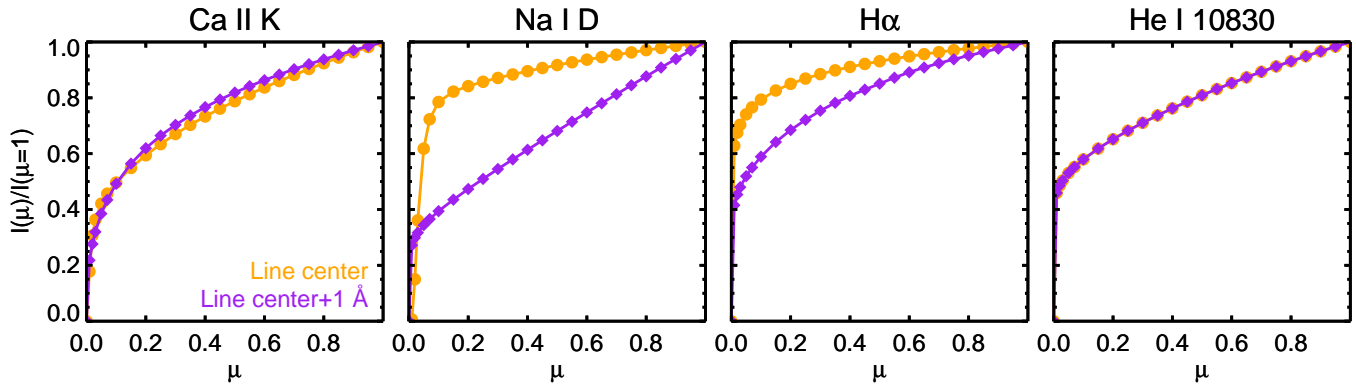}
   \figcaption{Normalized center-to-limb variations for $T_\text{eff} = 5500$ K. 
The line core is shown in orange and a $+1$ \AA\ offset from line center is shown in purple.
Note that the line cores of \ion{Na}{1} and H$\alpha$ tend to show less steep 
$I(\mu)$ functions compared with the line wing and continuum. \ion{He}{1} 10830
is not present in the photosphere of a 5500 K star and thus shows no difference
between the continuum at the two different wavelengths. These curves change
as a function of $T_\text{eff}$ with \ion{Na}{1} showing the most dramatic
differences.\label{fig:clvs}}
\end{figure*}

Accounting for the wavelength dependence of limb darkening across a single
spectral line is critical: center-to-limb variations, or CLVs, are
non-negligible for strong photospheric lines and can mimic planetary
atmospheric absorption \citep{yan15,czesla15,khalafinejad17,cauley17a,yan17}.
Examples of the intensity ratios for each line are shown in \autoref{fig:clvs}
for $T_\text{eff}=5500$ K. Note that CLVs are a function of $T_\text{eff}$ so
the curves in \autoref{fig:clvs} will change with different $T_\text{eff}$.

The \ion{Ca}{2} K curves behave differently than \ion{Na}{1} D and
H$\alpha$: the line core has a steeper limb-darkening function than the wing,
although the effect is small. This is also contrary to the solar observations
of \citet{white68} and \citet{ermolli10}. Both \citet{white68} and
\citet{ermolli10}, however, measure the wing limb-darkening at $\approx \pm 4$
\AA\ from line center, whereas we show the $\pm 1$ \AA\ case. The \ion{Ca}{2} K
line is much broader than \ion{Na}{1} D and H$\alpha$ for the $T_\text{eff}$ 
values we examine and thus the $\pm 1$ \AA\ case samples a steeper portion
of the line wing. The discrepency between our simulations and the
observed solar case can likely be attributed to such differences.

One important caveat regarding the CLVs calculated here is that
\texttt{SPECTRUM} is an LTE code. The cores of strong lines, such as H$\alpha$,
\ion{Na}{1} D, and \ion{Ca}{2} K, in main sequence G and K stars are formed
high up in the stellar photosphere, near the base of the chromosphere, where
LTE is no longer valid. Including non-LTE effects would produce more accurate
CLV curves and line profiles. However, as we will show, the dominant effects in
our simulations are the active region emission and coverage fraction
parameters. Thus any changes to the exact CLV values by including non-LTE
spectral synthesis will be small \citep[see Figure 11 of ][]{yan17} compared
with the effects of stellar active regions.

To make the simulations more computationally efficient, we do not
calculate a model spectrum at every $\mu$ value from disk center to the limb.
Instead, we use the derived limb darkening law to adjust the $\mu=1.0$ spectrum
at each disk position before that specific grid cell is added to the
disk-integrated spectrum. 

The facular contrast exhibits limb brightening relative to the underlying
photosphere, which is the result of the increased projected emitting area of
these optically thin features as the line-of-sight moves closer to the stellar
limb \citep{berger07}. This brightening is a function of wavelength, with bluer
wavelengths exhibiting larger effects. We adopt the solar limb brightening law
of \citet{gondoin08} (their equation 5).  While this treatment is not strictly
valid out to $\approx 10830$ \AA, where the average contrast is approximately
equal to the bolometric value, the influence on the simulations is much smaller
than other factors, e.g., the strength of the facular emission. We also
tested the limb brightening scaling from \citet{meunier10}, which is appropriate
for wavelengths between $4700$ \AA\ and $5500$ \AA, and found negligible
differences ($\lesssim 10\%$) to the \citet{gondoin08} scaling. Note that
the facular scaling is applied in addition to line core emission, i.e., even
at disc-center where the facular temperature difference is zero these is still
line emission. 

Facular limb brightening is a function of spectral type and magnetic field
strength \citep[e.g., ][]{beeck15}.  Thus a more precise approach would include
these effects by deriving brightening laws for individual spectral lines via
simulation.  Accounting for such effects is beyond the scope of this paper and
is likely unnecessary given the relatively small contribution of facular
brightening to the in-transit contrast spectra.  Facular contrast temperatures
are fairly similar to the photospheric temperature for the Sun \citep{unruh99}.
For this reason, and because $a_\text{act}$ and $f_\text{act}$ are the more
critical parameters, we fix the facular contrast temperature at $\Delta
T_\text{fac} = 50$ K.  The facular\footnote{We will use faculae to refer to all
bright active regions on the stellar surface and ignore the differences between
faculae, plage, and magnetic bright points.} spectra are limb brightened before
being added to the underlying photospheric spectrum, which is limb darkened. 

To simulate transits of active stellar surfaces, we consider the effects of
facular regions, spots, and the photosphere. We do not include filaments here
since our previous work found them to be unimportant relative to spots and
faculae \citep{cauley17a}. All spots and faculae are assumed to be circular and
facular regions are allowed to overlap to create more complicated shapes.
The radius of individual facular features ranges from a minimum of
three grid pixels, or 0.002\% of the stellar disk area, up to $\approx 10$ grid
pixels. Although faculae on the Sun can have sizes as small as a few hundred
kilometers \citep[e.g.,][]{title92,viticchie11}, or $\approx 10^{-7}$ of the
solar disk, the value of $(R_\text{p}/R_*)^2$ and the total facular coverage
fraction, combined with their positions on the disk, are much stronger
contibutors to the in-transit contrast effect. We thus do not consider smaller
facular sizes, as this only serves to add computational time to the
simulations. Features are placed randomly on the stellar disk with constraints
provided for different scenarios, e.g., active belts centered at a specific
latitude. Active regions and spots near the limb are foreshortened. 

We use the same model stellar spectra for the spots and photosphere, e.g., for
a star with $T_\text{eff} =5500$ K and a spot temperature difference of 500 K,
a $T_\text{eff} = 5000$ K model spectrum is selected for the spot. The relative
flux of the spot spectrum is scaled by the ratio of the model continuum intensities
of the $\mu=1.0$ (disk-center) spectra. For \ion{Ca}{2} K and \ion{He}{1} 10830
\AA\ we add an additional component to the spot spectrum: an emission line for
\ion{Ca}{2} K and an absorption line for \ion{He}{1} 10830 \AA\ (see
\autoref{fig:suncomps}). We fix the maximum emission value to 0.2 and the
minimum absorption to 0.8 for \ion{Ca}{2} K and \ion{He}{1} 10830 \AA,
respectively. The spot spectra are limb darkened using the derived
values for the appropriate spot temperature. 

The facular spectra are constructed by adding or subtracting a Gaussian to the
underlying photospheric spectrum depending on which line is being tested. The
strength of the intrinsic active region emission  is defined as $a_\text{act}$
and is the ratio of the Gaussian emission line to the photospheric absorption
line. Thus the same value of $a_\text{act}$ for different lines can produce
different levels of absolute flux depending on the depth of the photospheric
absorption; the ratio of the lines, however, is the same for identical values
of $a_\text{act}$. The FWHM of the Gaussian is fixed to 20 km s$^{-1}$, roughly
the width of the transmission spectra observed in
\citet{cauley15,cauley16,cauley17a}. We do not attempt to reproduce the exact
shape of the active region emission, e.g., the double-peaked \ion{Ca}{2} K
lines, since most observations cannot fully resolve these features. 

The contrast, or $a_\text{act}$ for our simulations, in bright active
regions shows a strong dependence on magnetic field strength for the Sun,
decreasing for field strengths up to $\approx 50$ G and then monotonically
increasing out to $\approx 1000$ G where the dependence plateaus 
\citep{kahil17}. Thus stars will most likely exhibit active regions with
varying levels of contrast across the disk. Since we are interested in the
average effect of these features on a planetary transmission signature, we
assume that all bright active regions have the same contrast. Letting
$a_\text{act}$ vary over the stellar disk could more strongly affect time
series absorption values, which we do not present here. 

\autoref{fig:suncomps} demonstrates that \ion{He}{1} 10830 \AA\ shows
absorption in active regions on the Sun. \citet{andretta95} find that
chromospheres $\approx 100 \times$ denser than the solar case can produce
\ion{He}{1} 10830 \AA\ in emission. We do not simulate these high-density cases
but we speculate that the results would be similar to the emission line cases
(H$\alpha$, \ion{Na}{1} D, and \ion{Ca}{2} K) explored here. In the case of
\ion{He}{1} 10830 \AA, $a_\text{act}$ refers to the depth of the Gaussian
absorption feature below the continuum, which is set equal to 1.0 in all
simulations.

Definitions and value ranges for the various parameters used in the simulations
are give in \autoref{tab:simpars}. We are motivated to search across a broad
parameter space in order to explore the extent to which observed exoplanet
transmission signals can be reproduced by the contrast effect. We find some
guidance from the Sun (see \autoref{sec:sun}). \autoref{fig:suncomps} shows
observed, spatially-resolved sunspot (top row) and facular (bottom row) spectra
(see \autoref{sec:sun} for details). For example, in the \ion{Ca}{2} H faculae
panel in \autoref{fig:suncomps}, $a_\text{act} \approx 1.0$, for H$\alpha$
$a_\text{act} \approx 0.6$, and for \ion{Na}{1} D $a_\text{act} \approx 0.8$.
Thus we explore a range of $a_\text{act}$ consistent with the solar case but
which includes values that may be more indicative of more active stars, e.g.,
$a_\text{act}$ can be as high as 3.0 or 4.0 in M-class flares
\citep{johnskrull97,xu16}. 

Spots are generally described as being composed of an inner umbral region and a
surrounding penumbral annulus. The umbral region is cooler than the penumbra,
showing a greater temperature differential with the surrounding photosphere.
This average differential of both umbral and penumbral regions, or $\Delta
T_\text{sp}$, depends on $T_\text{eff}$, with $\Delta T_\text{sp}$ increasing
with larger $T_\text{eff}$ \citep[see Figure 7 of][]{berdyugina05}. According
to \citet{berdyugina05}, $\Delta T_\text{sp}$ varies from $\approx 1000$ K for
$T_\text{eff} \approx 4500$ K up to $\approx 1800$ K for $T_\text{eff} \approx
5800$ K; there is little data for stars with $T_\text{eff} \gtrsim 5800$ K.  We
do not differentiate between umbral and penumbral regions, which can have
temperatures that differ by $\approx 1000$ K and corresponding differences
in the brightness of chromospheric line emission; the spot temperatures in the
simulations are considered an average of the two.  Thus we explore a different
range of $\Delta T_\text{sp}$ for each $T_\text{eff}$, varying from $\Delta
T_\text{sp} = 500$ K for $T_\text{eff} = 4500$ K up to $\Delta T_\text{sp} =
1500$ K for $T_\text{eff} = 6000$ K.  
 
Finally, spot and facular coverage fractions are chosen to simulate reasonable
G and K-star values. Sunspot disk coverage fractions vary from $\approx 0.001 -
0.010$ at solar minimum and maximum, respectively \citep{shapiro14}.  Spot
coverage fractions have been measured for active transiting exoplanet hosts:
\citet{morris17} find a mean disk coverage fraction of $3^{+6}_{-1}\%$ for
HAT-P-11 b; \citet{pont13} determine an average spot coverage fraction of
$\approx 1-2\%$ for HD 189733 b. Thus we conservatively explore values ranging
from 0.3\% up to the extreme case of 10\%.  For faculae, coverage fractions on
the Sun vary between $\approx 0.005$ and $0.04$ \citep{shapiro14}. Plage
coverage is similar, varying between $\approx 0.002$ and 0.06 \citep{mandal17}.
Combining the facular and plage coverages for the Sun, bright active regions
can account for up to $\approx 10\%$ of the solar disk. Estimating active
region coverage fractions for main-sequence G and K stars other than the Sun is
difficult. For this reason we extend our parameter exploration up to
$f_\text{act} = 50\%$ to account for the most active exoplanet hosts.  We note
that \citet{andretta17} find \ion{He}{1} 10830 \AA\ filling fractions as high
as $\approx 80\%$ for G and K-stars, although this is within the context
of their specific chromospheric models.

\begin{deluxetable*}{lcc}
\tablewidth{1.99\textwidth}
\tablecaption{Contrast model parameters and explored values\label{tab:simpars}}
\tablehead{\colhead{Parameter description}&\colhead{Symbol}&\colhead{Value range}}
\colnumbers
\tabletypesize{\scriptsize}
\startdata
Spot coverage fraction & $f_{sp}$ & 0.003-0.100  \\
Faculae coverage fraction & $f_\text{act}$ & 0.05 - 0.50  \\
Ratio of facular to photosphere core & $a_\text{act}$ & 0.5-3.0  \\
FWHM of active region emission/absorption & $\sigma_\text{act}$ & 20 km s$^{-1}$  \\
Central latitude of active region distribution & $\theta_\text{act}$ & $\pm 20^\circ-45^\circ$ \\
Temperature difference between spots and photosphere & $\Delta T_\text{sp}$ & 500 - 1500 K \\
Temperature difference between faculae and photosphere & $\Delta T_\text{fac}$ & 50 K \\
Minimum spot radius& $r_\text{sp}^\text{min}$ & 0.1-0.5 $R_\text{p}$ \\
Maximum spot radius & $r_\text{sp}^\text{max}$ & 0.3-1.0 $R_\text{p}$ \\
Planet-to-star ratio & ($R_\text{p} / R_*$)$^2$ & 0.1,0.2 \\
\enddata
\end{deluxetable*}

\section{Simulation results}
\label{sec:results}

Here we describe the results of the transit simulations for a variety of cases:
transits of the resolved solar disk (\autoref{sec:sun}), only star spots
(\autoref{sec:spots}), a uniform distribution of active regions
(\autoref{sec:uniform}), transiting active latitudes (\autoref{sec:actlats}),
and transiting slightly offset from active latitudes (\autoref{sec:offlat}).
The solar disk transits are useful for comparisons to solar analogs with
transiting hot planets and provide a realistic baseline example for the
simulations. The spot-only case is unlikely to occur but is useful for
demonstrating the contribution of spots to the contrast spectra. Finally,
transits of simulated surfaces with different bright active region geometries
cover the general (uniform active region distribution) and extreme (centered on
active latitudes) cases for which the contrast effect is important. 

\subsection{The Sun as a test case}
\label{sec:sun}

The Sun is the only star for which we have high-resolution spectra at high
spatial resolution of the stellar disk and thus provides a baseline example for
realistic active region parameters in the spectral lines explored here. Most
importantly, we are concerned with the \textit{intrinsic} emission and
absorption in spots and faculae. For stars with unresolved stellar surfaces,
there is a degeneracy between the strength of the active region spectral
feature and the coverage fraction of similar types of active regions \citep[see
Figure 17 of][]{cauley17a}. It is critical to break this degeneracy when
considering the contrast effect for transiting planets. There is some evidence
that the critical difference between active and less-active stellar spectra is
the coverage fraction and not the intrinsic emission strength:
\citet{andretta17} find a relatively narrow range of chromospheric densities
are required to reproduce the equivalent width ratios of \ion{He}{1} 5876 \AA\
and \ion{He}{1} 10830 \AA\ for G and K-type stars exhibiting a large range of
coverage fractions. More estimates of filling fractions for active stars are
needed to confirm this behavior.

\begin{figure*}[htbp]
   \centering
   \includegraphics[scale=.8,clip,trim=25mm 45mm 5mm 45mm,angle=0]{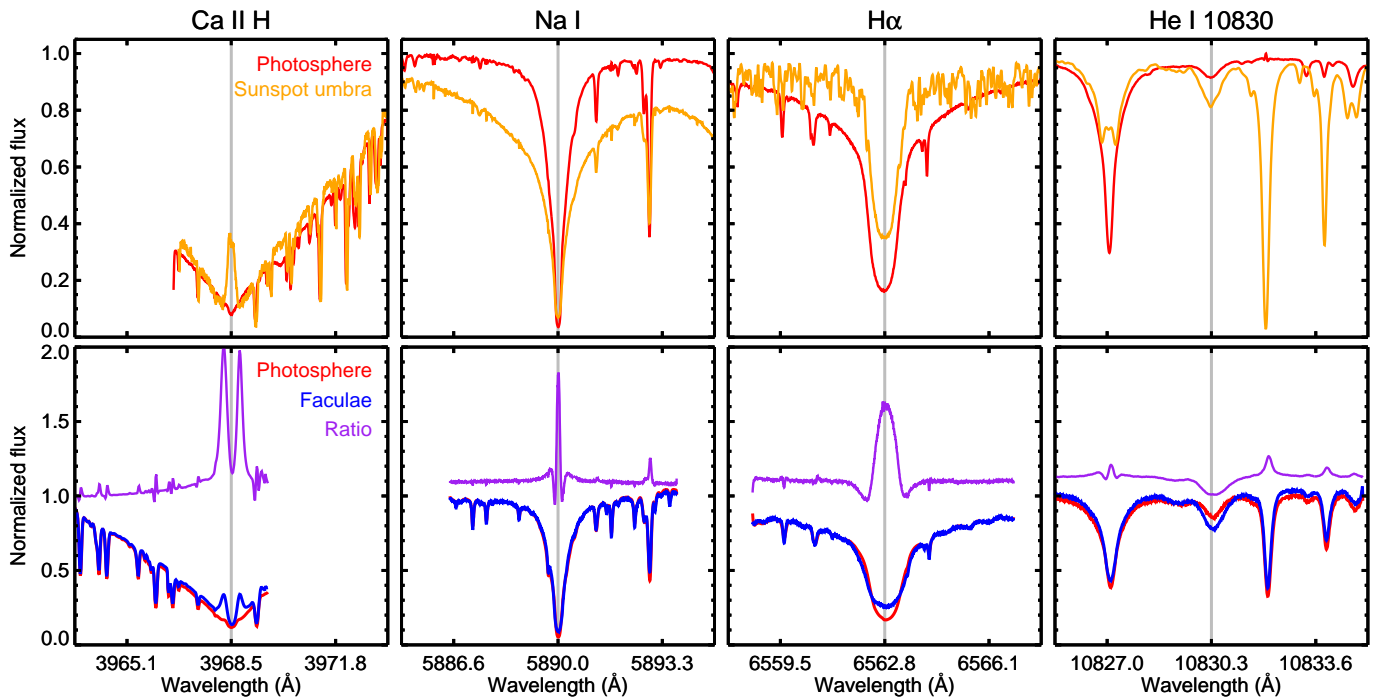} 
   \figcaption{Observed solar sunspot (upper row) and faculae (bottom row) spectra. In the upper panel,
   the spot spectra are normalized to highlight the line core contrast. The faculae-to-photosphere
   line ratio is shown in purple in the bottom row. The ratio is offset by 0.1 in the \ion{Na}{1}, H$\alpha$,
   and \ion{He}{1} panels. Ratios of $\approx 1.5 - 2.0$ are seen in \ion{Ca}{2} H, \ion{Na}{1}, and
   H$\alpha$. \ion{He}{1} 10830 \AA\ shows absorption relative to the photosphere due to
   the increased population of \ion{He}{1} atoms in the ground state of the 10830 \AA\ 
   transition, the result of the higher temperature in facular regions. \label{fig:suncomps}}
\end{figure*}

In order to provide some context for intrinsic active region features in our
simulations, we have collected observed spectra of spots and facular regions on
the Sun. These spectra are shown in \autoref{fig:suncomps}. The data are from a
variety of sources. The sunspot umbra and quiet photosphere spectra in the
upper row of \autoref{fig:suncomps} are from the National Solar Observatory
Solar Flux Atlas\footnote{http://diglib.nso.edu/flux}. We normalize both the
spot and quiet photosphere spectra to the continuum flux at $\pm 7$ \AA\ to
emphasize the differences in the line core. 

The plage and photosphere spectra for \ion{Na}{1} and H$\alpha$ in the bottom
row are from \citet{kuckein16}.  The faculae spectra are averages of $5\times5$
pixel boxes of the brightest regions in sections 5 and 7 from their Figure 2.
The quiet Sun photosphere spectrum is an average of all other pixels in
sections 1 through 10, excluding the filament and bright regions selected for
the facular spectra. 
 
The \ion{He}{1} 10830 \AA\ data was acquired in May 2013 with the Tenerife
Infrared Polarimeter \citep[TIP-II][]{collados07} attached to the Vacuum Tower
Telescope \citep{vonderluehe98}. The spectrum is an $10\times10$ pixel average
taken within an active region, avoiding the sunspot.  The \ion{Ca}{2} H faculae
spectra are extracted from the data presented in \citet[][see their Figure
3]{beck11}.  Note that we were not able to find suitable \ion{Ca}{2} K spectra
so \ion{Ca}{2} H spectra are used as an approximation.  We construct facular
spectra by producing a normalized core flux image and then selecting pixels
with fluxes $> 1.5$. We only use pixels with y-coordinate values $< 40''$ in
order to avoid the solar limb. Spectra from the selected pixels are averaged to
produce the facular spectrum in \autoref{fig:suncomps}.  The quiet Sun
photosphere spectrum is an average of the pixels with normalized flux values
$0.9 < F_\text{N} < 1.1$ in the same region. 

\autoref{fig:suncomps} shows that faculae-to-photosphere ratios in the line
cores, which is represented by $a_\text{act}$ in the simulations, are $\approx
1.5 - 2.0$ with the largest contrast in \ion{Ca}{2} H. The contrast is likely
$\approx 1.5 -2.0\times$ higher for \ion{Ca}{2} K. \ion{He}{1} shows the unique
behavior among these activity indicators of going further into absorption in
facular spectra. Note that the absorption seen in the \ion{He}{1} 10830 \AA\
photospheric spectrum is due to the diffuse overlying chromosphere, and is not
intrinsic to the photosphere, which could not be completely separated when 
constructing the line profiles. The enhanced facular absorption is a combination
of higher temperatures inducing more collisional excitation and enhanced
photoionization by coronal back illumination and subsequent recombination to
the lower level of the 10830~\AA\ transition \citep{andretta97}. The enhanced
lower level population leads to higher opacity and increased line absorption.
We note that for the densest chromospheres in \citet{andretta97} the 10830 \AA\
transition goes into emission. Energetic solar flares also show \ion{He}{1}
10830 \AA\ in emission \citep{judge15,kuckein15}. However, the increased
absorption in solar facular regions shown here is likely more representative of
the average 10830 \AA\ spectrum in the cases discussed in this paper.  

\begin{figure*}[htbp]
   \centering
   \includegraphics[scale=.7,clip,trim=8mm 45mm 5mm 60mm,angle=0]{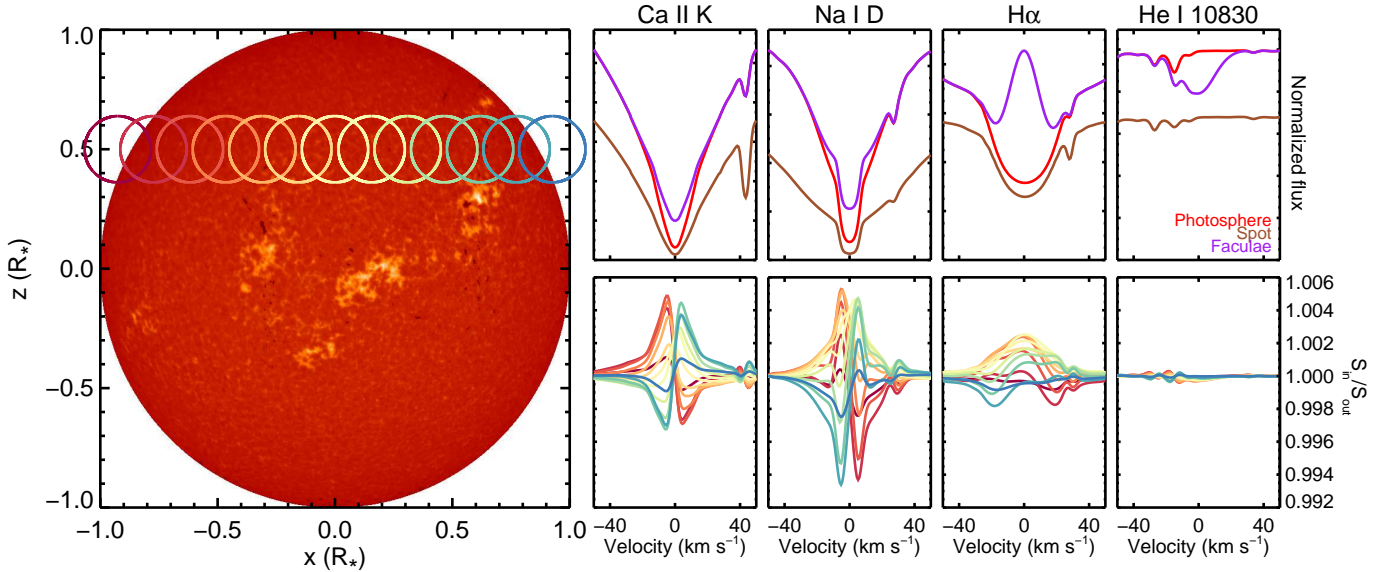}
   \figcaption{Examples of contrast spectra as a function of transit phase for the Sun
on 2016 July 14. The solar disk in the left panel is a narrowband \ion{Ca}{2} K image
rebinned to the spatial resolution (0.0025 $R_*$) used in the simulations. The solar spin
axis is in the positive vertical direction. Note
that the color scaling is not necessarily representative of the relative brightness
of the various features across the disk.
The disk shows $\approx 7\%$ facular coverage and $\approx 0.2\%$ spot coverage.  
The colored open circles show the specified planet positions, moving from red to blue
as the transit progresses. 
The upper-right panels show the normalized spectra for the various disk features and spectral lines.
The y-axis plot ranges are chosen to highlight the relative core emission strength;
the continuum is not necessarily shown.
Note that the weak lines blueward of \ion{He}{1} 10830 \AA\ are not associated 
with \ion{He}{1}.
The bottom-right panels show $S_\text{T} = S_\text{in}/S_\text{out}$ for the planet positions
indicated by the colored circles in the left panel. There is little contribution to $S_\text{T}$
from the active regions; the signal is almost entirely due to CLVs.
\label{fig:transun}}
\end{figure*}

\autoref{fig:transun} shows a transit example using $(R_\text{p}/R_*)^2=0.02$
of a narrowband \ion{Ca}{2} K solar disk snapshot from the Chromospheric
Telescope \citep[ChroTel;][]{bethge11} taken at 13:30:00.00 UT on 2016 July 14.
In order to identify spots and active regions in the image, we created a
normalized image and assigned points with normalized flux values $\geq 1.2$ as
faculae (bright regions in the image in \autoref{fig:transun}) and $\leq 0.8$
as spots (darkest regions in \autoref{fig:transun}, e.g., near the active
region in the upper-right quadrant). Note that the spot identification also
flagged regions that are more likely filaments.  These regions contribute
negligibly to the total flux. This snapshot shows $\approx 7\%$ facular
coverage and $\approx 0.2\%$ spot coverage, which corresponds to a moderate
level of solar activity. We set $a_\text{act} = 1.6$ for \ion{Ca}{2} K,
H$\alpha$, and \ion{Na}{1} D; for \ion{He}{1} 10830 \AA\ $a_\text{act} = -0.2$.
The photospheric and spot spectra are $T_\text{eff} = 5800$ K and $\Delta
T_\text{sp} = 500$, respectively. Note that $\Delta T_\text{sp} = 500$ is more
representative of the average brightness temperature of the combined umbral and
penumbral regions; the darkest umbral regions in sunspots typically have
temperatures $\approx 1000-1500$ K lower than the surrounding photosphere
\citep{sutterlin98}.

The bottom row of \autoref{fig:transun} shows the in-transit spectra for the
planet positions indicated by the colored circles in the image. For the most
part, the only contributions to the contrast spectrum
$S_\text{in}/S_\text{out}$ are CLVs; the active regions near disk-center do not
have much of an effect.  The solar example illustrated some important effects.
Primarily, stars with a low coverage fraction of active regions will not
exhibit a large contrast effect. Also of note is the absence of any contrast
signatures at \ion{He}{1} 10830 \AA; $a_\text{act}$ is smaller in this case
compared with the other lines and the contrast spectra are correspondingly
weaker. Despite the lack of contrast signatures in the solar case, we explore a
wide range of parameters in the following sections in order to understand the
possible extent of contrast contamination. 

\subsection{The negligble effect of starspots and filaments}
\label{sec:spots}

As noted in \citet{cauley17a}, spots tend to have a small effect on the
contrast spectra of the lines explored here. The primary reason for this is the
flux-weighted contribution of spots to the integrated spectrum: their cooler
temperature produces a smaller fraction of the total flux relative to the same
area of photosphere. Note that this is only the case for spot coverage
fractions of $\lesssim 15\%$. We do not consider larger coverage fractions,
such as those explored in \citet{rackham18}, since there is no evidence of
such large spot areas in main sequence G and K stars.  

\begin{figure*}[htbp]
   \centering
   \includegraphics[scale=.9,clip,trim=10mm 50mm 0mm 40mm,angle=0]{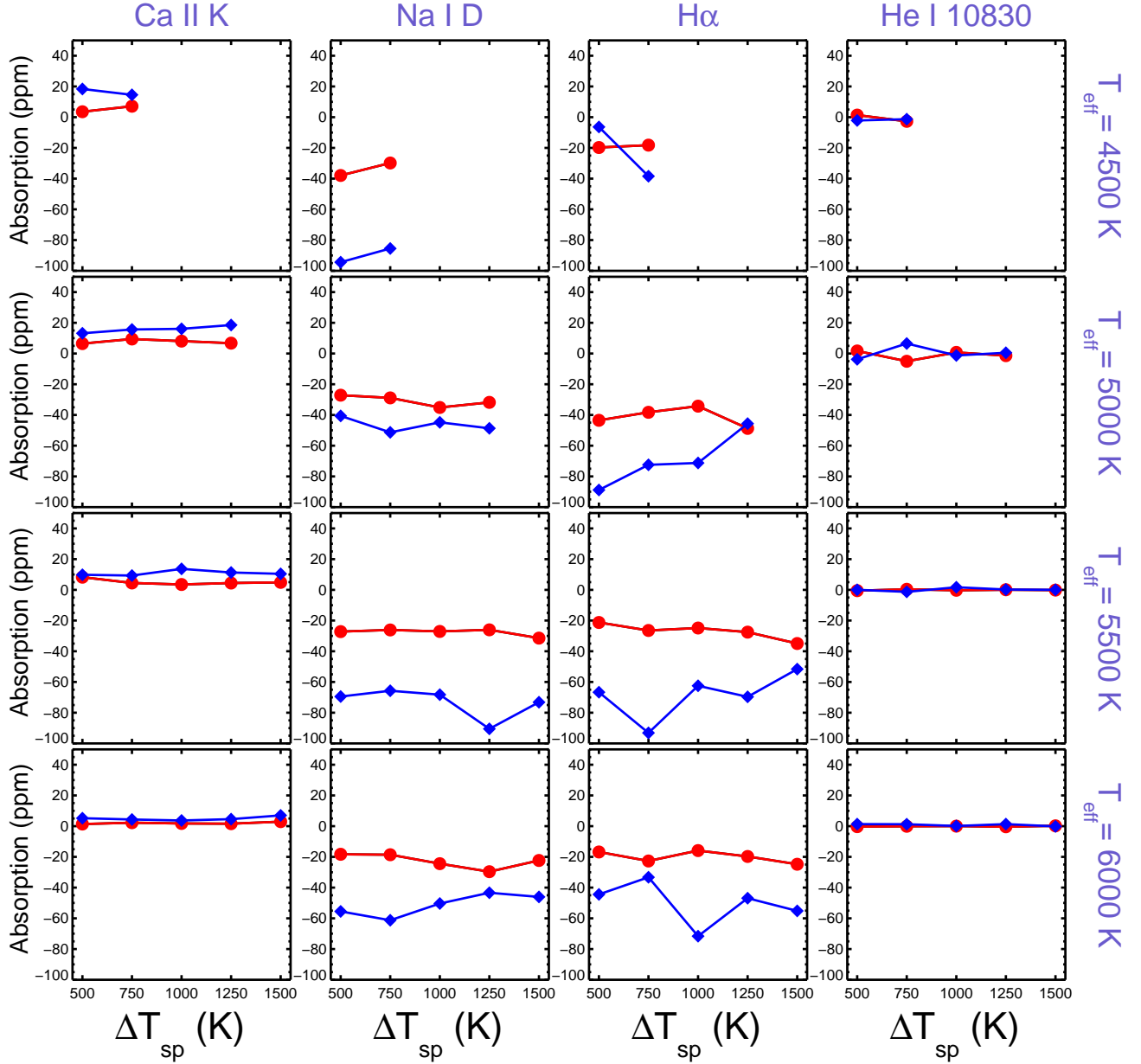}
   \figcaption{Mean in-transit absorption signatures for stellar surfaces
with $f_\text{sp}=10\%$ as a function of $\Delta T_\text{sp}$ (see \autoref{sec:model}
for details on the choice of $\Delta T_\text{sp}$). Columns are individual lines and
rows show the same $T_\text{eff}$. The red lines are for $(R_\text{p}/R_*)^2=0.01$ and
the blue lines represent $(R_\text{p}/R_*)^2=0.02$. The
absorption in the line is measured in parts per million (ppm) and all cases
are shown with the same absorption range of $-100$ to 50 ppm for comparison. None of
the cases shown exhibit any contrast effect above $\approx 0.01\%$, even for
this extreme example of a 10\% coverage fraction.
\label{fig:spots}}
\end{figure*}

\autoref{fig:spots} shows an example for stellar surfaces with a spot coverage
fraction of 10\%. The red lines show the $(R_\text{p}/R_*)^2=0.01$ case and the
blue lines show the 0.02 case. The absorption is measured across a 2
\AA\ band, centered on the rest wavelength of the line, of the mean
$S_\text{in}/S_\text{out}$ spectrum across the stellar disk. Note that the
absorption is in \textit{parts per million}. \autoref{fig:spots} demonstrates
how small the contrast effect is from spots for atomic lines, indepedent of
$\Delta T_\text{sp}$. While it is possible that transits of spot groups may
produce larger signals for a small portion of the transit, consistent
in-transit absorption signatures are unlikely to be due to star spots.
Furthermore, the strength of atomic absorption typically detectable in
high-resolution transmission spectra is $\approx 0.1 \%$, about $10-1000\times$
the strength of the contamination found here for the pure spot case. Thus it
not currently feasible to disentangle such small effects from the measurements.
The active region scenarios explored in the rest of the paper all include a
$0.3\%$ spot coverage fraction for completeness. 

Filaments are embedded in the chromosphere and corona and are made of cool
material when compared to their surroundings \citep{martin98}. Magnetic field
lines support them against gravity. They appear in absorption against the solar
disk but are seen as bright prominences when observed above the limb. We
explored the magnitude of the contrast effect due to filaments in
\citet{cauley17a} and found them to be unimportant compared to faculae and
spots. Thus we ignore them here and refer the reader to our previous work for
more details.

\subsection{Uniform distribution of active regions}
\label{sec:uniform}

\begin{figure*}[htbp]
   \centering
   \includegraphics[scale=.7,clip,trim=8mm 45mm 5mm 60mm,angle=0]{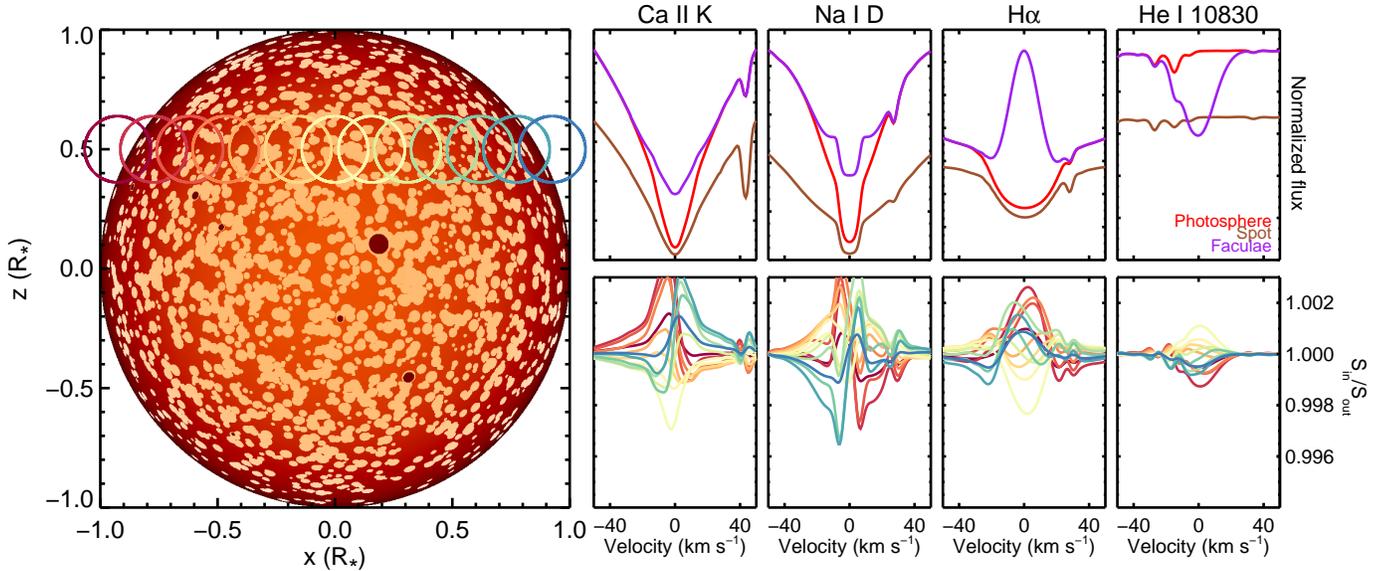}
   \figcaption{Same format as \autoref{fig:transun} except for a simulated uniform distribution of active regions. 
   Note that colors do not accurately represent the relative brightness of surface features
and are used for illustrative purposes.
   The example shows the case of $f_\text{act} = 0.4$, $a_\text{act} = 3.0$ (-0.4 for \ion{He}{1} 10830 \AA), 
   $(R_\text{p}/R_*)^2 = 0.02$, and $T_\text{eff} = 5500$ K. The left panel shows the stellar disk configuration, with
   light regions representing active regions, darker regions representing spots, and the open colored
   circles showing the planet positions at which contrast spectra are calculated. The upper-right
   panels show the normalized photospheric, spot, and faculae spectra for each line. The bottom-right
   panels show the contrast spectra $S_\text{in}/S_\text{out}$ as a function of transit phase, where
   the colors map to those in the stellar disk image. The strongest effect can be seen mid-transit where
   the contrast spectra can reach depths of $\approx 0.5\%$ and absorption values of $\approx 0.02\%$.
\label{fig:tranuni}}
\end{figure*}

One broad category of active region coverage is a uniform distribution across
the stellar disk. \autoref{fig:tranuni} shows a $(R_\text{p}/R_*)^2 = 0.02$
transit example for $f_\text{act} = 0.4$, the active region coverage fraction,
and $a_\text{act} = 3.0$ ($-0.4$ for \ion{He}{1} 10830 \AA), the flux relative
to that in the line core of the photospheric spectrum. Note that the
\ion{He}{1} 10830 \AA\ contrast spectra show \textit{emission}, the opposite of
the other lines. We do not show the emission and absorption components of the
lines for spot spectra (see \autoref{fig:suncomps} for the solar case); only
the scaled model spectrum of the appropriate spot $T_\text{eff}$ is given for
clarity.

\begin{figure*}[htbp]
   \centering
   \includegraphics[scale=.9,clip,trim=10mm 50mm 0mm 30mm,angle=0]{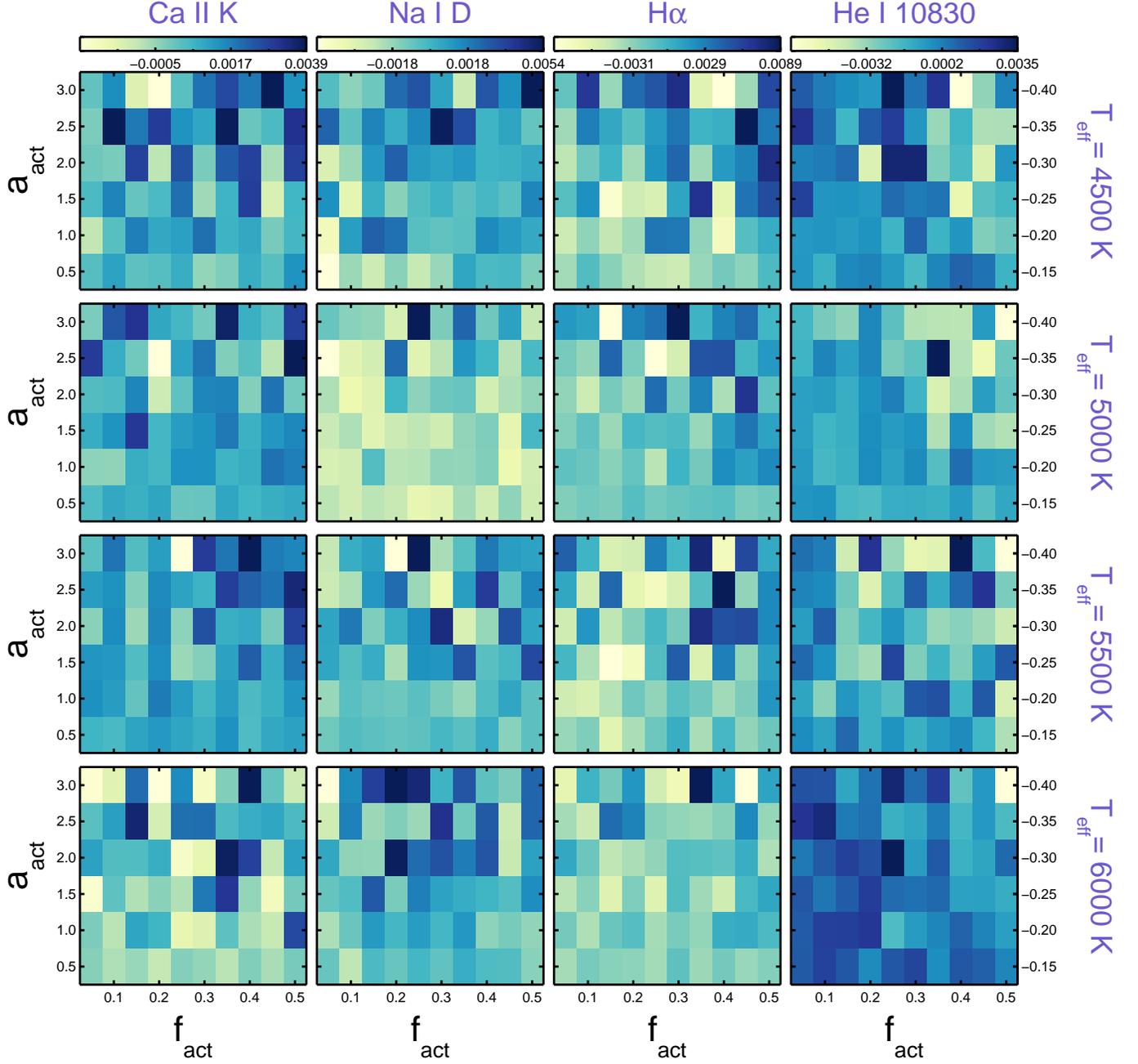}
   \figcaption{Mean in-transit absorption for transits of a stellar disk with a uniform
distribution of active regions for
the case of $(R_\text{p}/R_*)^2 = 0.01$. Columns are individual lines and rows
are $T_\text{eff}$. The x-axis $f_\text{act}$ is the active region coverage fraction
of the stellar disk and the y-axis $a_\text{act}$ is the strength of the
active region emission. Note that the \ion{He}{1} 10830 \AA\ vertical
scale is different from the other lines. The values indicated by the color
bars are in percent absorption (integrated over $\lambda_0 \pm 1$ \AA). All
panels are on the same absolute color scale. The effects are almost entirely negligible
for most lines, suggesting that even large active region coverage fractions
cannot produce strong contrast signals if they uniformly distributed.
\label{fig:uniform01}}
\end{figure*}

\begin{figure*}[htbp]
   \centering
   \includegraphics[scale=.9,clip,trim=10mm 50mm 0mm 30mm,angle=0]{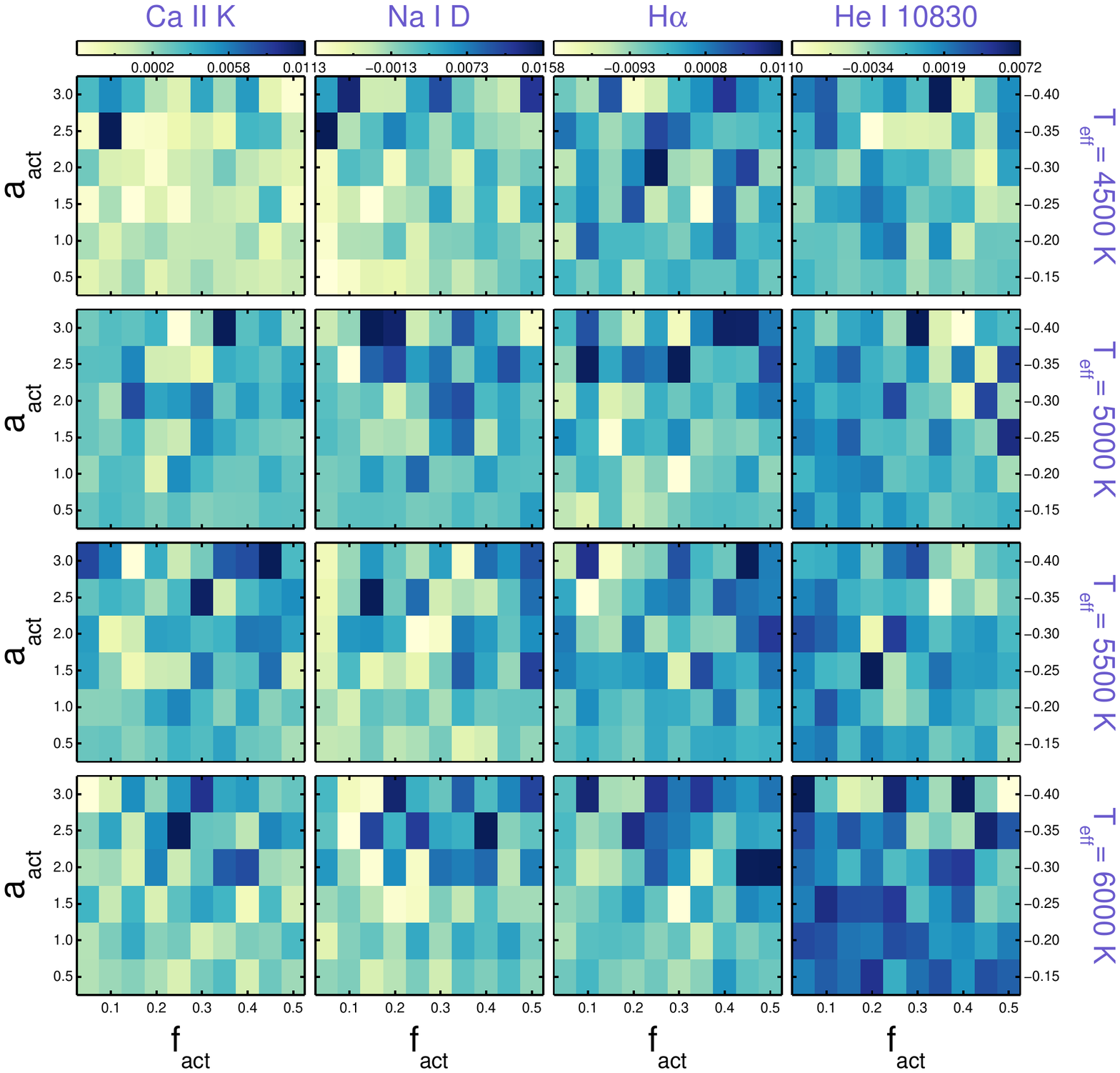}
   \figcaption{Same as \autoref{fig:uniform01} but for a uniform distribution of active regions for
the case of $(R_\text{p}/R_*)^2 = 0.02$. Similar to the $(R_\text{p}/R_*)^2 = 0.01$ case, the effects are small for all
lines, never reaching values greater than $\approx 0.006\%$.
\label{fig:uniform02}}
\end{figure*}

\autoref{fig:uniform01} and \autoref{fig:uniform02} show absorption maps for
the $(R_\text{p}/R_*)^2 = 0.01$ and $(R_\text{p}/R_*)^2 = 0.02$ cases,
respectively.  The color bars show the absorption, in percent, measured
for the mean $S_\text{in}/S_\text{out}$ spectrum sampled across the transit.
The strength of the contrast absorption roughly increases as $a_\text{act}$ and
$f_\text{act}$ increase, i.e., from the lower-left to the upper-right of each
map. This can be seen most clearly for the cases of \ion{Na}{1} D with
$T_\text{eff} = 4500$ K and H$\alpha$ with $T_\text{eff} = 5500$ K.  However,
absorption from the contrast effect is $\lesssim 0.005\%$ for all lines and the
difference between even the largest and smallest values is $\approx 0.005\%$. 
The weak absorption in the uniform case is dominated by noise, with
small variations in the location of active regions producing random
changes in the measured absorption. This suggests that a consistent
in-transit absorption signal measured for a transiting exoplanet
is unlikely to be caused by a uniform active region distribution.

\subsection{Active latitudes}
\label{sec:actlats}

In \citet{cauley17a} we showed that the only scenario capable of producing the
observed in-transit H$\alpha$ transmission spectrum for HD 189733 b was the
transit of a very active latitude. To produce a consistent in-transit
absorption signature, the planet must continuously transit active regions with
similar activity levels, i.e., chromospheric emission or absorption.
Here we extend the active latitude analysis to all of the lines and
a broader range of active region parameters. 

\begin{figure*}[htbp]
   \centering
   \includegraphics[scale=.7,clip,trim=8mm 45mm 5mm 60mm,angle=0]{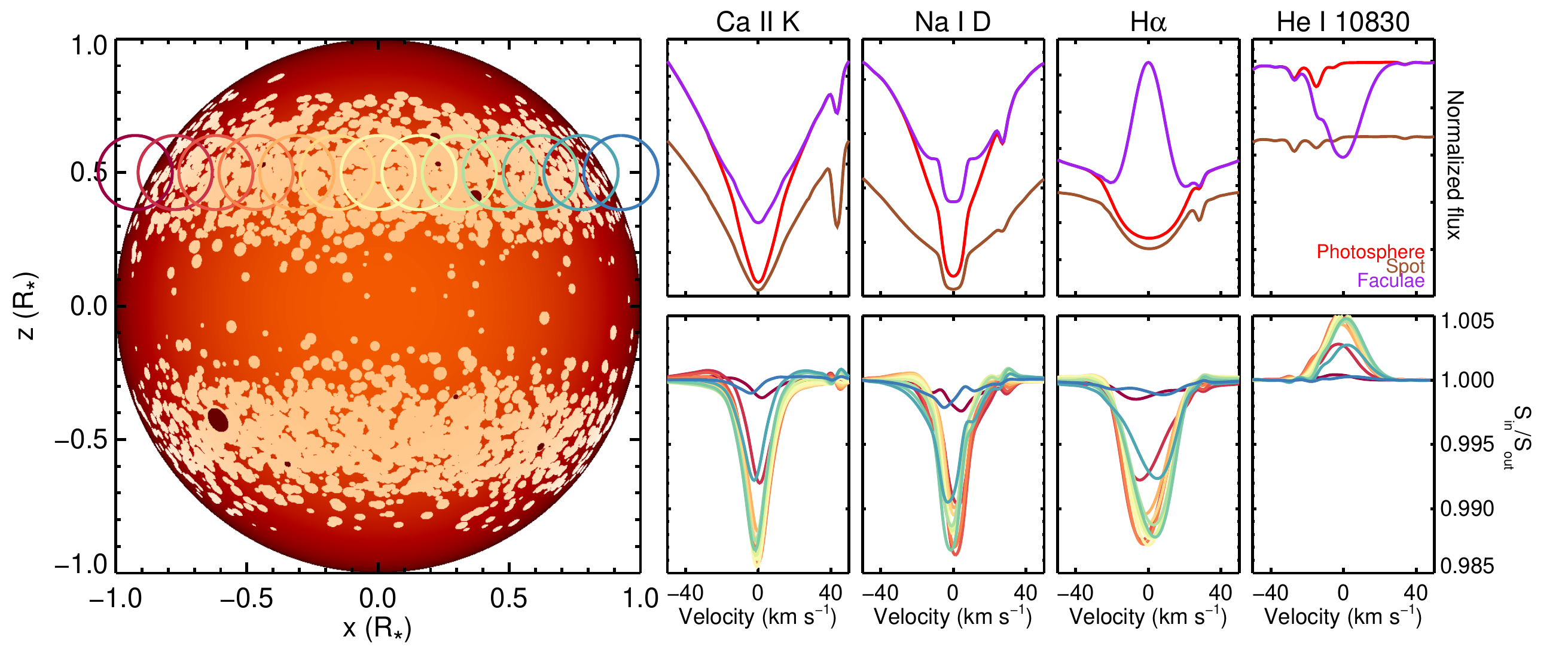}
   \figcaption{Examples of contrast spectra as a function of transit phase for an active latitude case.
   The parameters used and the format is the same as \autoref{fig:tranuni} except the active
   regions are distributed normally around the planet's transit latitude. The contrast effect is
   much stronger due to the constant occultation of faculae by the planet.
\label{fig:tranact}}
\end{figure*}

\autoref{fig:tranact} shows an example of an active latitude transit for a
transit chord centered on the active latitude. This is critical: the average
contrast spectrum depends strongly on whether or not the planet transits within
$\approx \pm 5^\circ$ of the active latitude center.  Off-latitude transits are
explored in \autoref{sec:offlat}. It is clear from \autoref{fig:tranact} that
the contrast spectra are much stronger than in the uniform distribution case
(see \autoref{fig:uniform01} and \autoref{fig:uniform02}).

One noteworthy feature in \autoref{fig:tranact} is the behavior of \ion{He}{1}
10830~\AA: because this line is in absorption in most normal stellar
chromospheres, transits of active regions seen at 10830 \AA\ will show
\textit{emission} instead of absorption. This suggests that any \ion{He}{1}
10830 \AA\ atmospheric absorption signatures will tend to be masked by the
contrast effect, not enhanced. This is encouraging for \ion{He}{1} 10830 \AA\
transmission spectra since detections will likely be lower limits and
can confidently be attributed to the planet's atmosphere and not to
stellar active regions.  

\begin{figure*}[htbp]
   \centering
   \includegraphics[scale=.9,clip,trim=10mm 50mm 0mm 30mm,angle=0]{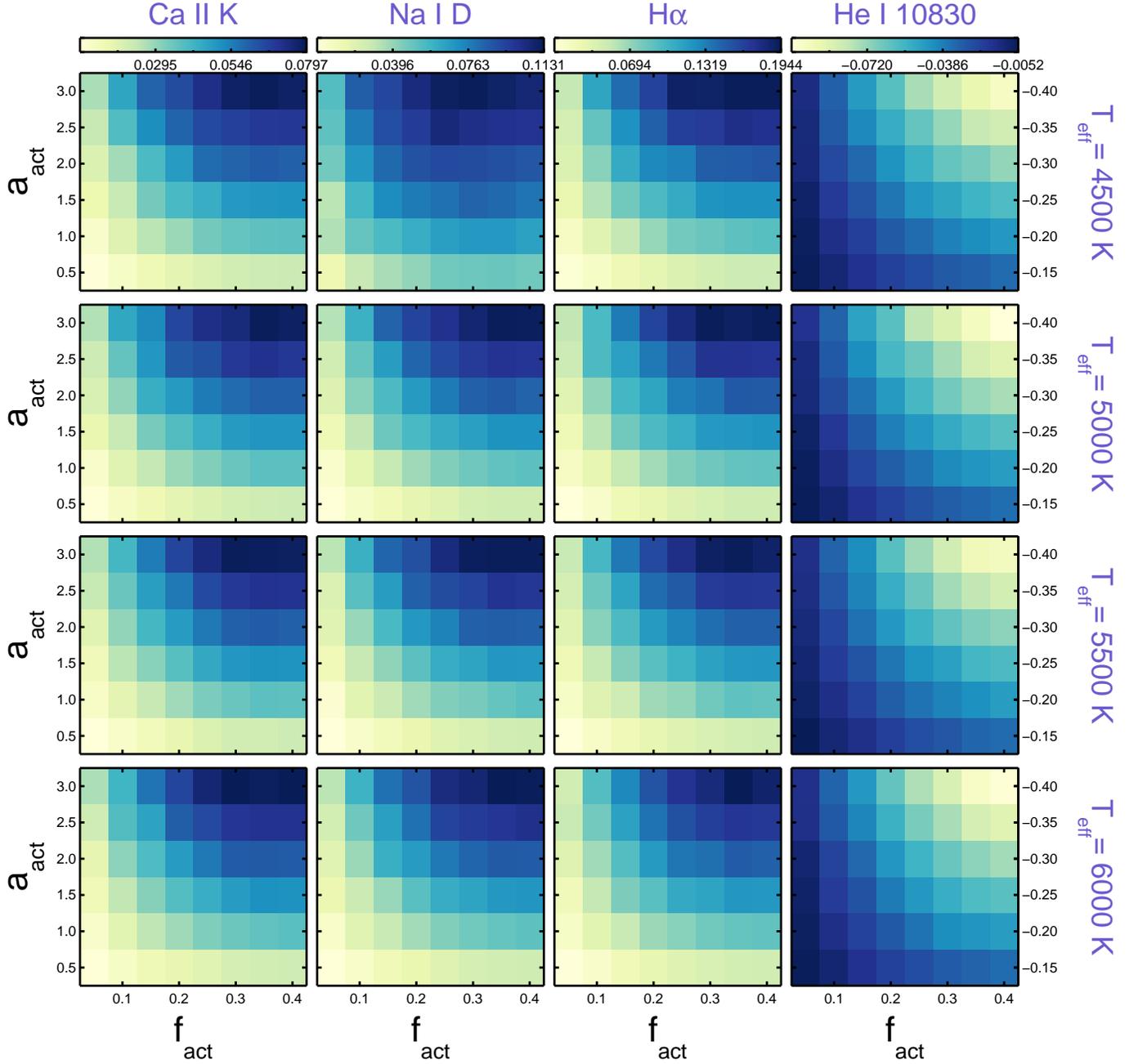}
   \figcaption{Example of absorption maps for active latitude transits for
the case of $(R_\text{p}/R_*)^2 = 0.01$. All formatting is the same as \autoref{fig:uniform01}. 
The effects are much larger than the uniform cases (\autoref{fig:uniform01} and \autoref{fig:uniform02}), 
although still small for most lines.
The contamination values for \ion{Na}{1} D approach $0.1\%$, a non-negligible
fraction of the measured absorption in some exoplanet atmospheres. 
\label{fig:actlat01}}
\end{figure*}

\begin{figure*}[htbp]
   \centering
   \includegraphics[scale=.9,clip,trim=10mm 50mm 0mm 30mm,angle=0]{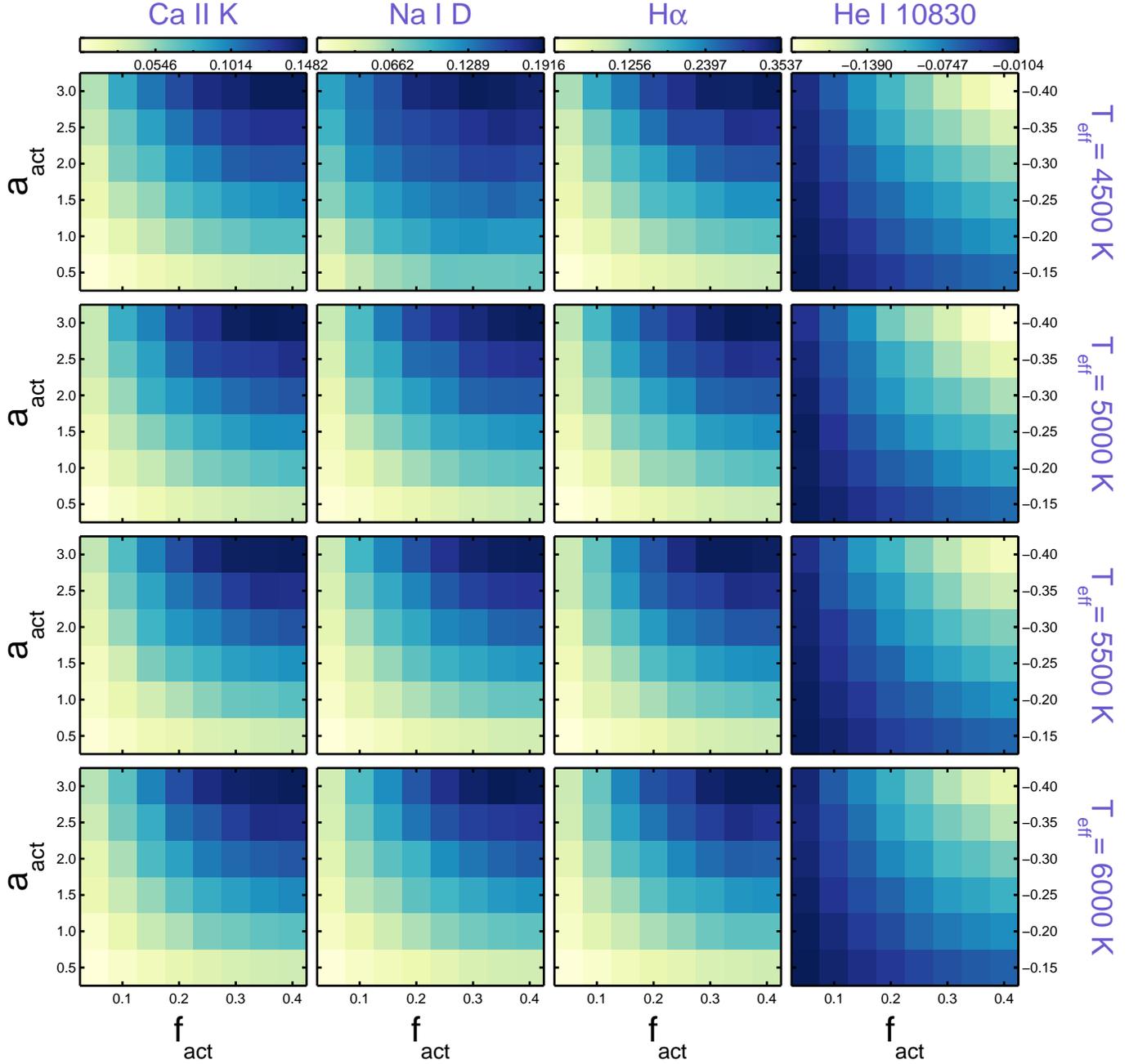}
   \figcaption{Example of absorption maps for active latitude transits for
the case of $(R_\text{p}/R_*)^2 = 0.02$. All formatting is the same as \autoref{fig:uniform01}. 
The magnitude of the contrast contamination is $\approx 2\times$ the amount
seen in the $(R_\text{p}/R_*)^2 = 0.01$ case and approaches observed values for the most
extreme cases (upper-right corners of each map; see \autoref{tab:obsvals} for
observed values of H$\alpha$ and \ion{Na}{1} D). 
\label{fig:actlat02}}
\end{figure*}

\autoref{fig:actlat01} and \autoref{fig:actlat02} show the absorption maps for
active latitude transits. The absorption increases approximately linearly with
$a_\text{act}$ and $f_\text{act}$. There is little variation with
$T_\text{eff}$, suggesting that the constrast effect is not strongly affected
by the underlying photospheric spectrum but rather the strength and
distribution of active regions. The strongest absorption signatures
are seen in the $(R_\text{p}/R_*)^2 = 0.02$ case for H$\alpha$ and \ion{Na}{1} D,
which are $\approx 0.35\%$ and $\approx 0.18\%$, respectively. Comparisons
of the contrast simulation absorption to observed values will be
discussed in \autoref{sec:discussion}.  

\begin{figure}[htbp]
   \centering
   \includegraphics[scale=.4,clip,trim=35mm 25mm 15mm 30mm,angle=0]{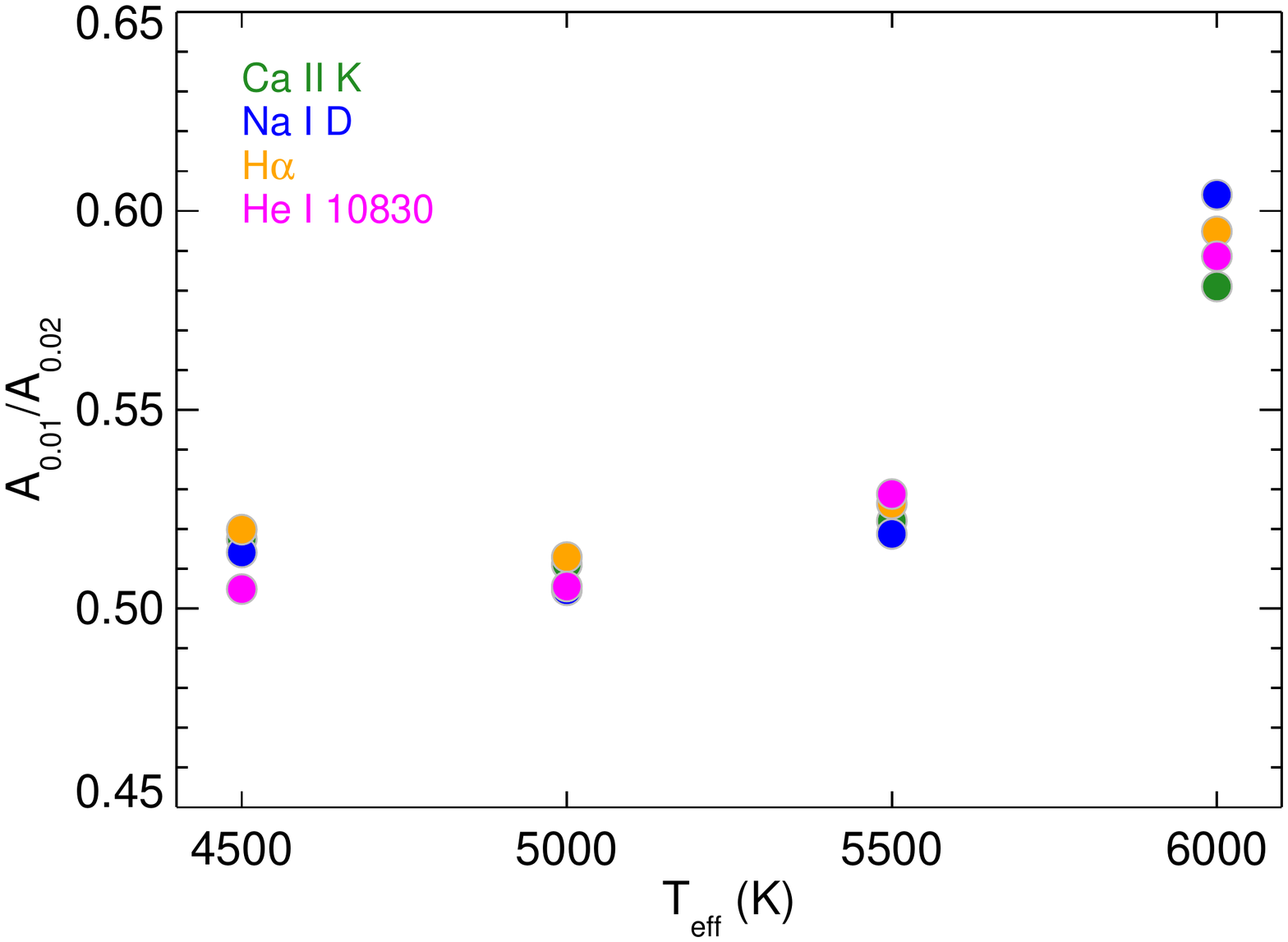}
   \figcaption{Absorption ratio between $(R_\text{p}/R_*)^2=0.01$ and 0.02 for 
the active latitude case of $a_\text{act} = 3.0$ and $f_\text{act} = 0.4$. The
ratio is not exactly 0.5 and tends to increase with increasing $T_\text{eff}$.
\label{fig:absrat}}
\end{figure}

We note that the absorption in the $(R_\text{p}/R_*)^2 = 0.01$ case is not
necessarily 50\% of that in the $(R_\text{p}/R_*)^2 = 0.02$ case.
\autoref{fig:absrat} shows the absorption ratio for the
$(R_\text{p}/R_*)^2=0.01$ and 0.02 simulations in the case of $a_\text{act} =
3.0$ and $f_\text{act} = 0.4$.  The ratio is not exactly 0.5 and tends to
increase with $T_\text{eff}$. Although we have not been able to isolate the
cause of this trend, it is likely due to differences in limb darkening curves
as a function of $T_\text{eff}$.

\subsection{Transiting away from active latitudes}
\label{sec:offlat}

We showed in \autoref{sec:actlats} that transits of active latitudes can
produce significant contrast signatures in certain cases. However, those cases
were for active regions distributed normally around the planet's transit
latitude. This increases the likelihood that the planet covers active regions
and produces a contrast signal. Here we explore how the contrast effect changes
as the active region latitude moves away from the planet's transit latitude. 

\begin{figure*}[htbp]
   \centering
   \includegraphics[scale=.7,clip,trim=8mm 45mm 5mm 60mm,angle=0]{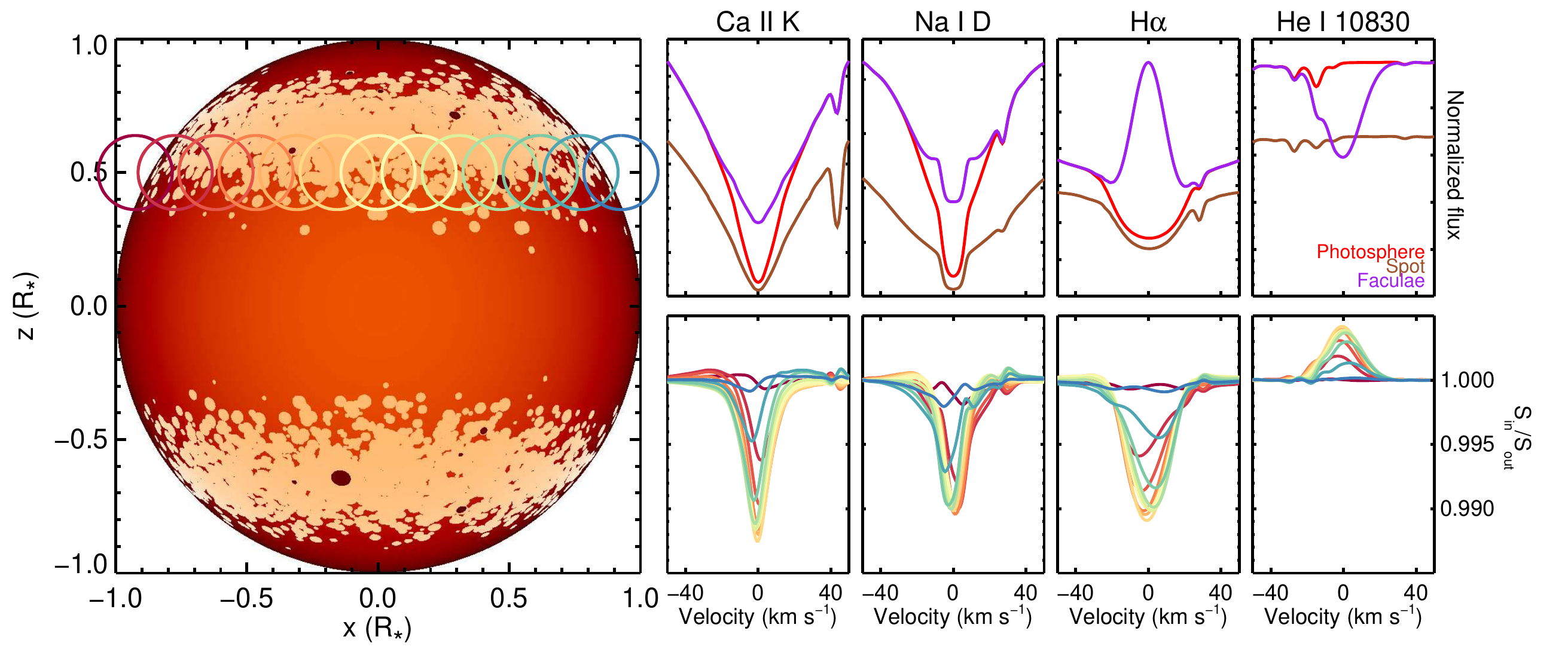}
   \figcaption{Examples of contrast spectra as a function of transit phase for
a transit $\approx 10^\circ$ off the center of an active latitude.  The
parameters used and the format is the same as \autoref{fig:tranuni}. The active
latitudes are at $\pm 40^\circ$. The contrast effect is weaker compared with
the active latitude case in \autoref{fig:actlat02} since the planet transits
portions of the disk with a lower density of active regions. 
\label{fig:tranoff}}
\end{figure*}

\begin{figure*}[htbp]
   \centering
   \includegraphics[scale=.7,clip,trim=8mm 45mm 5mm 90mm,angle=0]{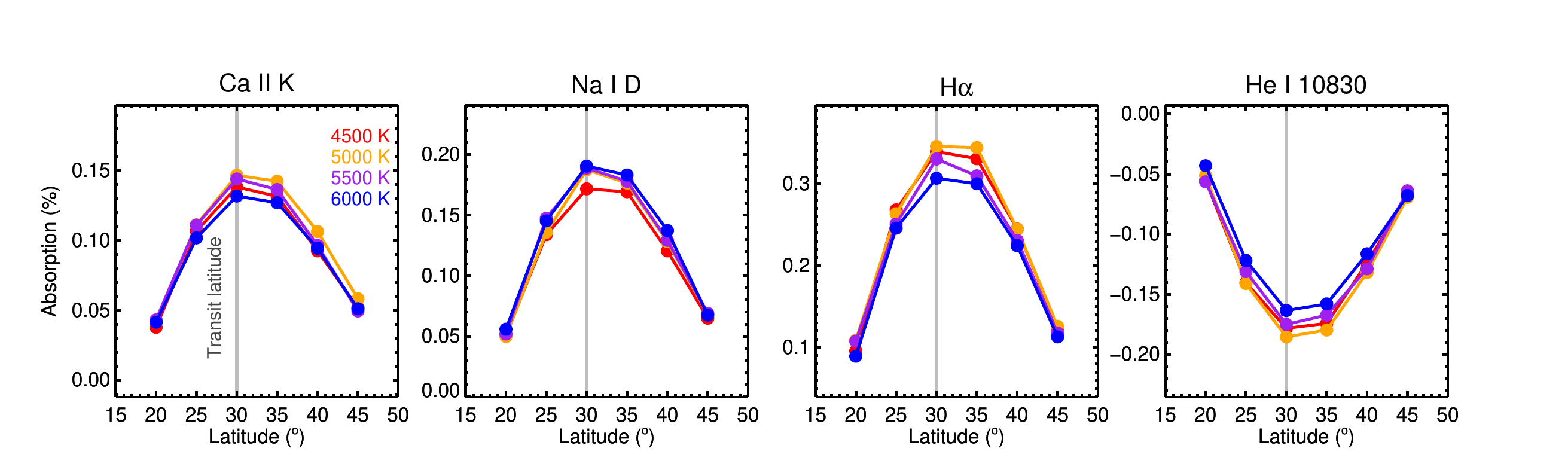}
   \figcaption{Contrast absorption as a function of central latitude for
active latitudes. Colors represent $T_\text{eff}$ and
the planet's transit latitude is shown with the vertical gray line. Note that
more negative numbers imply a stronger effect for \ion{He}{1} 10830 \AA. The
contrast absorption decreases sharply when the active latitudes are centered
away from the planet's transit latitude. 
\label{fig:offlat}}
\end{figure*}

\autoref{fig:tranoff} shows an example of an off-latitude transit and
\autoref{fig:offlat} shows the contrast absorption as a function of active
region latitude with a planet transit latitude of 30$^\circ$ ($b = 0.5$) and
$(R_\text{p}/R_*)^2=0.02$. For this example we fix the active region coverage
fraction and facular core ratio to $f_\text{fac} = 0.35$ and $a_\text{act} =
3.0$, respectively. For \ion{He}{1} 10830~\AA\ we use $a_\text{act} = -0.4$.
These values were chosen because they produce strong contrast spectra in the
active latitude transit case.

For each spectral line the contrast absorption decreases sharply as the active
region latitude moves away from the planet's transit latitude. This suggests
that in order to produce significant contrast signals active latitudes need to
be centered within $\leq 5^\circ$ from the planet's transit latitude. The need
for the planet to transit very close to the active latitude argues against
consistent in-transit absorption signatures being produced by the contrast
effect. For this to be the case, the active latitude must cover almost the
entire circumference of the star at the planet's transit latitude. Furthermore,
the active belt cannot migrate away from the transit chord, which is unlikely
if the star experiences activity cycles comparable to the Sun. The simulations
presented here thus confirm the result in \citet{cauley17a} for the in-transit
H$\alpha$ signal observed for HD 189733 b. 

\section{Discussion}
\label{sec:discussion}

Our simulations show that the contrast effect for chromospheric lines in the
optical can be non-negligible for certain active region configurations and
intrinsic emission strengths. Randomly distributed active regions, however,
even in cases of coverage fractions of $\approx 50\%$, cannot produce
strong contrast spectra.  Active latitudes centered $\leq 5^\circ$ from the
planet's transit latitude are necessary to create contrast absorption lines
with central depths of $\approx 1\%$. 

How do the strongest contrast scenarios compare with observations?
\autoref{tab:obsvals} lists the current published absorption detections for
\ion{Na}{1} D and H$\alpha$. We do not include comparisons for \ion{Ca}{2} K
and since there are no reported detections for \ion{Ca}{2} K in the literature
that can be attributed to atmospheric absorption\footnote{See \citet{cauley17a}
for possible in-transit \ion{Ca}{2} H and K absorption signatures}.
Furthermore, the absorption cores of these lines are most strongly affected by
NLTE effects, which we do not account for. The only \ion{He}{1} 10830 \AA\
detection \citep{spake18} is given its own discussion below. The bottom of the
table lists the \ion{Na}{1} 5890 \AA\ and H$\alpha$ values obtained for the
case of $(R_\text{p}/R_*)^2=0.02$ and $T_\text{eff} = 5500$ K and $a_\text{act}
= 3.0$. The value of $f_\text{act}$ is given in brackets.  

Even the extreme case of $f_\text{act} = 0.5$ for uniformly distributed active
regions cannot account for $10\%$ of the absorption in any of the
published values. Similarly, off-latitude transits can only produce up to
$\approx 50\%$ of the observed values. Contrast absorption for active latitude
transits, however, approaches the published values for both \ion{Na}{1} D and
H$\alpha$. Absorption as strong as $\approx 1\%$, such as that detected
in \ion{Na}{1} D for WASP-49 b \citep{wyttenbach17} and WASP-69 b \citep{casasayas17}
 is still well above the maximum values found in our simulations. 

While our simulations show that it may be possible to produce contrast
signatures similar to what is observed in HD 189733 b, as suggested by
\citet{barnes16}, we argue that this is unlikely. As we have seen, in order to
produce the strongest contrast signatures the planet must be transiting very
close to the central latitude of an active band on the stellar surface. In
addition, the strength of the intrinsic emission in the active regions must be
$\approx 2\times$ that seen in solar active regions. Finally, this same
scenario must persist across many epochs in order to consistently produce
in-transit absorption, such as is repeatedly observed for HD 189733 b in both
H$\alpha$ and \ion{Na}{1} D
\citep{redfield08,jensen11,wyttenbach15,khalafinejad17,cauley17a}.  The
specificity of the necessary active region parameters argues against the
contrast effect as the mechanism responsible for the transmission absorption
measured for planets transiting active stars. Nonetheless, a small fraction of
the measured absorption is likely due to the transiting of stellar active
regions. This should be taken into consideration when considering discussions
of H$\alpha$ transmission spectra. 
 
\begin{deluxetable*}{lccccc}
\tablewidth{1.99\textwidth}
\tablecaption{Observed atomic transmission spectra absorption\label{tab:obsvals}}
\tablehead{\colhead{}&\colhead{}&\colhead{$R$}&\colhead{Integration bin}&
\colhead{\ion{Na}{1} D \AA}&\colhead{H$\alpha$}\\
\colhead{Reference}&\colhead{Object}&\colhead{($\lambda / \Delta \lambda$)}&\colhead{(\AA)}&
\colhead{(\%)}&\colhead{(\%)}}
\colnumbers
\tabletypesize{\scriptsize}
\startdata
\citet{charbonneau02} & HD 209458 b & 5540 & 1.2 & 0.023 & \nodata \\
\citet{snellen08} & HD 209458 b & 45,000 & 1.5 & 0.07 & \nodata \\
\citet{redfield08} & HD 189733 b & 60,000 & 12.0 & 0.067 & \nodata \\
\citet{jensen11} & HD 189733 b & 60,000 & 12.0 & 0.053 & \nodata \\
\citet{jensen12} & HD 189733 b & 60,000 & 6.0 & \nodata & 0.302 \\
\citet{cauley15} & HD 189733 b & 68,000 & 2.0 & \nodata & 0.293 \\
\citet{wyttenbach15} & HD 189733 b & 115,000 & 1.5 &  0.141 & \nodata \\
\citet{cauley16} & HD 189733 b & 68,000 & 2.0 & 0.082 & 0.685 \\
\citet{khalafinejad17} & HD 189733 b & 60,000 & 1.5 & 0.340 & \nodata \\
\citet{wyttenbach17} & WASP-49 b & 115,000 & 0.4 & 1.440 & \nodata  \\
\citet{chen17} & WASP-52 b & 1100 & 16.0 & 0.378 & \nodata \\
\citet{casasayas17} & WASP-69 b & 115,000 & 1.5 & 5.8 & \nodata \\
\citet{casasayas18} & KELT-20 b & 115,000 & 0.75 & 0.178 & 0.594 \\
 & & & & & \\
Uniform [$f_\text{act} = 0.50$] & $(R_\text{p}/R_*)^2 = 0.02$ & 70,000 & 2.0 & -0.003 & $< 0.001$\\
Active latitude [$f_\text{act} = 0.40$] & $(R_\text{p}/R_*)^2 = 0.02$ & 70,000 & 2.0 & 0.183 & 0.325\\
Off-latitude [$f_\text{act} = 0.35$]& $(R_\text{p}/R_*)^2 = 0.02$ & 70,000 & 2.0 & 0.059 & 0.111\\
\enddata
\end{deluxetable*}

The usefulness of \ion{He}{1} 10830 \AA\ as a diagnostic of exoplanetary
exospheres was recently reignited by \citet{oklopcic18} \citep[see also
][]{seager00,turner16}. Soon after, the first detection of \ion{He}{1} 10830
\AA\ in an exoplanet atmosphere was reported by \citet{spake18} for WASP-107 b.
Our simulations suggest that \ion{He}{1} 10830 \AA\ transmission absorption can
be contaminated at the 0.1\% level for transits of active latitudes.  However,
any exosphere absorption should be \textit{decreased} by the planet's transit
of active regions since \ion{He}{1} 10830 \AA\ is mostly in absorption in G and K
star active regions. The only exception to this would be for very dense
chromospheres which show \ion{He}{1} 10830 \AA\ in emission, an unlikely scenario
even for main-sequence M-dwarfs \citep{andretta97}. 

The relative absorption signature measured by \citet{spake18} is 0.049\%
integrated over 98 \AA.  A similar integration bin for our simulated
\ion{He}{1} 10830 \AA\ contrast spectra, in the case of an active latitude
transit for $f_\text{act} = 0.40$, $a_\text{act} = -0.40$, $T_\text{eff} =
4500$ K, and $(R_\text{p}/R_*)^2 = 0.02$, gives an emission contribution to the
contrast spectrum of $\approx -0.004\%$.  This is $\approx 10\times$ lower than
the magnitude of the WASP-107 b measurement and is in emission instead of
absorption. Thus the signal from \citet{spake18} may be diluted at the $\approx
10\%$ level but, as \citet{spake18} also discuss, cannot be the cause of transiting
a quiet photospheric chord.  Due to the small contrast effect in the line,
\ion{He}{1} 10830 \AA\ should be considered favorable compared with H$\alpha$
or \ion{Na}{1} D as a transmission spectroscopy tool for probing extended
exoplanet atmospheres. 

H$\alpha$ remains an under-utilized tool for probing the thermospheres of hot
planets. The presence of $n = 2$ hydrogen in hot planet atmospheres is
well-established theoretically \citep{christie13,huang17}. Due to its ease of
access with echelle spectrographs, the somewhat higher signal-to-noise in the
line core compared with \ion{Na}{1} D, and the fact that it is relatively free
of telluric contamination, combined with our findings that even active stars
most likely cannot produce strong contrast spectra, H$\alpha$ should be
considered in future transmission spectroscopy studies of hot planets.  This is
especially the case for extremely hot planets such as KELT-9 b \citep{gaudi17}
and KELT-20 b \citep{lund17} where the temperature is high enough for collisional
excitation of hydrogen to become important \citep{huang17}. 

\section{Conclusion}
\label{sec:conclusion}

We have explored how stellar activity in the form of spots and bright facular
or plage regions can contribute to the high-resolution optical transmission
spectra in chromospherically active lines for giant planets. Overall, the
emission from stellar active regions in the simulated lines has a weak effect
on the transmission spectrum and varies little with $T_\text{eff}$. Only
specific geometries, where the active region distribution is within $\approx
5^\circ$ of the planet's transit chord, combined with coverage fractions
$\gtrsim 20\%$ and active region emission strengths $2-3\times$ the
photospheric line strength can produce signatures similar to those observed for
certain hot planet systems. In particular, observed H$\alpha$ and \ion{Na}{1} D
absorption can most likely be attributed to the planet's atmosphere in the case
of HD 189733 b, although some of the signal probably originates
in stellar active regions, especially in the case of H$\alpha$. 

\ion{He}{1} 10830 \AA\ is a promising exosphere diagnostic \citep{spake18} and
we find that atmospheric absorption should be depressed relative to the true
values due to \ion{He}{1} 10830 \AA\ being in absorption in stellar
chromospheres. However, the effect is on the order of $\approx 0.1\%$ even for
active region coverage fractions of $\approx 0.4$, suggesting that stellar activity
should be unimportant for exoplanets with moderate predicted \ion{He}{1} 10830
\AA\ absorption \citep[e.g., GJ 436 b and HD 209458 b;][]{oklopcic18}. This
strengthens the case for the utility of the 10830 \AA\ line as a diagnostic
of exoplanet atmospheres.  

The era of extremely large telescopes (ELTs) will enable high-resolution
transmission spectra of super-Earths and small rocky planets.  The effects of
planets transiting active regions scales with the value of
$(R_\text{p}/R_*)^2$, suggesting that the relative magnitude of the contrast
effect should be similar for super-Earths and rocky planets. However,
transmission spectrum light curves will likely exhibit higher levels of
temporal variability since the relative size of the planet to spots and active
regions decreases, increasing the frequency with which the planet transits 
individual active regions. Simulations of the contrast effect for super-Earths
and rocky planets should be pursued.

While we have attempted to explore a broad parameter space for giant planet
transits of active stellar surfaces, we caution against specific comparisons of
exoplanet systems with absorption values derived here.  More precise knowledge
of the active region distributions and emission strength for exoplanet host
stars is needed to reach firmer conclusions about the absolute contribution of
active regions to exoplanet transmission spectra. Magnetic mapping and modeling
of faculae and spot contributions to the spectra and brightness variations of
planet hosting stars will be useful in this respect
\citep[e.g.,][]{dumusque14,herrero16,fares17}. Non-LTE effects should also be
included in order to produce more precise estimates of the contrast effect in
specific spectral lines. 

\bigskip

{\bf Acknowledgments:} We thank the anonymous referee for their comments and
suggestions, which helped improve the manuscript. This work is supported by
NASA Origins of the Solar System grant No. NNX13AH79G (PI: E.L.S.). A portion
of this work is also supported by the National Science Foundation through
Astronomy and Astrophysics Research Grant AST-1313268 (PI: S.R.). CD, CK, and
MV were supported by grant DE 787/5-1 of the Deutsche Forschungsgemeinschaft
(DFG). This work has made use of NASA's Astrophysics Data System. Vacuum Tower
Telescope in Tenerife and ChroTel are operated by the Kiepenheuer-Institute for
Solar Physics, Freiburg, Germany, at the Spanish Observatorio del Teide, of the
Instituto de Astrof{\'i}sica de Canarias. The ChroTel filtergraph has been
developed by the Kiepenheuer-Institute in co-operation with the High Altitude
Observatory in Boulder, CO, USA.

\software{SPECTRUM, \citet{gray94},
http://www.appstate.edu/~grayro/spectrum/spectrum.html}

\end{document}